\documentclass[journal]{IEEEtran}

\usepackage{amsmath}	
\usepackage{graphicx}	
\usepackage{amssymb}
\usepackage{amsfonts}
\usepackage{makecell}
\usepackage{color}
\usepackage{cite}
\usepackage{algpseudocode}  
\usepackage{hyperref}
\usepackage{authblk}
\makeatletter  
\newif\if@restonecol  
\makeatother

\usepackage[linesnumbered,ruled,vlined]{algorithm2e}

\begin{document}
\title{Evaluate PAC codes via Efficient Estimation on Weight Distribution}
\author{Junhua You,
        Shaohua Wu,
        Yajing Deng
        and Qinyu Zhang 
 \thanks{J. You, S. Wu, Y. Deng and Q. Zhang are with the School of Electronics and Information Engineering, Harbin Institue of Technology (Shenzhen), Guangdong, China (e-mail:youjunhua@stu.hit.edu.com; hitwush@hit.edu.cn; hitdengyj@163.com; zqy@hit.edu.cn).}
}
\maketitle
\begin{abstract}
  In this letter, we introduce an efficient method for estimating weight distributions of polar codes and polarization-adjusted convolutional (PAC) codes. Based on a recursive algorithm of computing the weight enumerating functions of polar cosets, this method focuses on two key objectives: accurately determining the number of low-weight codewords and quickly approximating the distribution of high-weight codewords. Simulation results demonstrate that this hybrid method maintains competitively low complexity while effectively achieving the objectives.
\end{abstract}

\IEEEpeerreviewmaketitle

\section{Introduction}
\IEEEPARstart{P}{olar} codes have garnered substantial attention over the preceding decade and have been incorporated into the standard of the fifth-generation (5G) mobile communication technology as the pioneering channel coding technique capable of achieving the symmetric (Shannon) capacity of a binary-input discrete memoryless channel (BI-DMC). However, notwithstanding their considerable advantages over alternative channel coding schemes, polar codes still exhibit the limitation of a relatively small minimum distance. To overcome this limitation, several precoding techniques have been introduced, including CRC (Cyclic Redundancy Check)-polar codes which are widely used in the present. Among these precoding techniques, polarization-adjusted convolutional (PAC) codes, as detailed in the seminal work \cite{arikanSequentialDecodingChannel2019} by Arikan, have recently gained prominence due to their remarkable capability to reach the finite-length capacity bound.

The key rationale behind the superior performance of PAC codes compared to traditional polar codes lies in the convolutional precoding's effective enhancement of the weight distribution. Indeed, gaining a deep understanding of these weight distributions holds significant implications for the continued refinement of both PAC and polar codes. Nevertheless, it's imperative to acknowledge that computing the weight distribution of channel codes, which is an NP-hard problem, remains a formidable and intricate endeavor.
 
Exhaustive enumeration is a commonly employed technique for obtaining the weight distributions of short linear block codes. However, its computational complexity grows exponentially with the data length, rendering it unviable for longer codes. If we narrow our focus to the enumeration of low-weight codewords, a commonly used method, as described in \cite{liAdaptiveSuccessiveCancellation2012}, is to transmit an all-zero codeword through a low-noise channel into a successive cancellation list (SCL) decoder with a large list size to capture low-weight codewords. Some probability-based estimation methods \cite{valipourProbabilisticWeightDistribution2013,zhangEnhancedProbabilisticComputation2017,seyedmasoumian2022approximate,li2023weight} have been proposed that can relatively quickly estimate the numbers of low-weight codewords but cannot guarantee their accuracy. Besides, some deterministic recursive algorithms have also been proposed. \cite{miloslavskayaComputingPartialWeight2022} introduced a recursive decomposition method that simplifies the problem of enumerating low-weight codewords for the entire code into similar problems for subcodes. In two separate works \cite{niu2019polar,polyanskayaWeightDistributionsSuccessive2020}, algorithms that compute the weight distribution of polar cosets along the all-zero decoding path are proposed. In \cite{yaoDeterministicAlgorithmComputing2023}, an efficient methodology for computing the weight distribution of polar cosets along any arbitrary decoding path is proposed, enabling the precise computation of the weight distribution for a (128,64) 5G polar code within two hours. Additionally, closed-form expressions for enumerating low-weight codewords are explored in \cite{rowshanClosedformWeightEnumeration2023}. These endeavors have yielded significant progress; however, in practical applications, spending several hours or even days to obtain the weight distribution can be frustrating. Therefore, a swift and effective approach to obtaining code weight distributions is a crucial task.

The inspiration for our work is derived from \cite{liAdaptiveSuccessiveCancellation2012} and \cite{yaoDeterministicAlgorithmComputing2023}. In this work, we propose an efficient method to estimate the whole weight distribution for polar codes and PAC codes. This method is rooted in the concept of polar cosets, enabling us to obtain the distributions of both low-weight and high-weight codewords by the weight enumeration functions of a select subset of polar cosets.
The paper is organized as follows. The polar channel coding technique is reviewed in Section \ref{sec2}. The proposed method to estimate the weight distribution is introduced in Section \ref{sec3}. Section \ref{sec4} provides the simulations and analysis. Finally, the paper is concluded in Section \ref{sec5}.

\section{Preliminaries}
\label{sec2}
\subsection{Brief Review on Polar Codes and PAC Codes}
Assuming $N=2^n$, a $(N,k)$ polar code is generated by selecting $k$ rows from the polar matrix $F_N=B_NK_2^{\otimes m}$. Here, $B_N$ represents the bit-reversal permutation matrix, and $K_2^{\otimes m}$ denotes the $m$-th Kronecker power of $
  K_2=\left[\begin{array}{ll}
    1 & 0 \\
    1 & 1
    \end{array}\right].
$

The encoding of polar codes is given by $\mathbf{x}=\mathbf{u}F_N$, where $\mathbf{u}$ is a length-$N$ binary input vector carrying $k$ data bits, and $\mathbf{x}$ is the codeword for transmission. The positions of the $k$ data bits in $\mathbf{u}$ are specified by an information index set $\mathcal{A}$ of size $k$, with $\mathcal{A}\subseteq \{0,1,\cdots,N-1\}$. The remaining $N-k$ bits in $\mathbf{u}$ are set to $0$, which are called \emph{frozen bits}. We also use $\mathcal{F}= \{0,1,\cdots,N-1\}\setminus \mathcal{A}$ to denote the frozen index set that specifies the positions of the frozen bits. For the traditional polar codes, the generation matrix $G_N=F_N$.

The generator matrix of the PAC code has been modified to $G_N = TF_N$, where $T$ is a convolution operation which is characterized by an impulse response $\mathbf{c}=(c_0,c_1,\cdots,c_m)$ assuming that $c_0 \neq 0$ and $c_m \neq 0$. The pre-transforming matrix $T$ can be represented in an upper-triangular Toeplitz matrix form.

\subsection{Polar Coset and Its Weight Enumerating Function}
For a vector $\mathbf{u}_i\in \{0,1\}^{i+1}$ with $0\leq i \leq N-1$, we define the \emph{polar coset} for path $\mathbf{u}_i$ as the affine space
\begin{equation*}
  C_{N}\left(\mathbf{u}_{i}\right) \triangleq \left\{\left(\mathbf{u}_{i}, \mathbf{u}^{\prime}\right) G_{N} \mid \mathbf{u}^{\prime} \in\{0,1\}^{N-i-1}\right\}
\end{equation*}
where $\left(\mathbf{u}_{i}, \mathbf{u}^{\prime}\right)$ represents the concatenation of $\mathbf{u}$ and $\mathbf{u}^\prime$. The weight enumerating function for polar coset $C_{N}\left(\mathbf{u}_{i}\right)$ as the polynomial is defined as 
\begin{equation*}
  {A_{N}\left(\mathbf{u}_{i}\right)(X) \triangleq  \sum_{w=0}^{N} A_{w} X^{w}}
\end{equation*}
where $A_{w}$ is the number of vectors in $C_{N}\left(\mathbf{u}_{i}\right)$ with Hamming weight $w$.

\cite{yaoDeterministicAlgorithmComputing2023} proved that any polar codes can be the union of disjoint polar cosets and proposed a deterministic recursive algorithm to compute the weight enumerating function of polar cosets, which is indicated by {\tt{CalcA($N$,$\mathbf{u}_{i-1}$)}} in the following
 sections.

\section{Efficient Estimation on Weight Distribution}
\label{sec3}

In this section, we present a hybrid method aimed at the efficient estimation of weight distributions for polar codes and PAC codes. This approach is rooted in the concept of polar cosets, enabling us to obtain the distributions of both low-weight and high-weight codewords by the weight enumeration functions associated with a select subset of polar cosets.

\subsection{Number of Low-Weight Codewords}
\label{sec_3_subsec_1}
The number of low-weight codewords plays a crucial role in characterizing the block error rate (BLER) performance of block codes. On one hand, the minimum weight of codewords directly corresponds to the minimum distance of the block code. On the other hand, the number of low-weight codewords dominates the union bound of the block codes. For polar codes, SCL decoders with a large list size are often employed to capture low-weight codewords\cite{liAdaptiveSuccessiveCancellation2012}. While this technique is accurate and efficient in determining the count of minimum-weight codewords, it poses challenges when estimating the count of other low-weight codewords, often necessitating even larger list sizes and consequently escalating computational complexity. In this part, we enhance this technique by polar cosets.

First we define the \emph{last frozen index} of a polar code $\mathbb{C}$ as $
  {\tau(\mathbb{C}) \triangleq \max \{\mathcal{F}\}}
$
and the \emph{mixing factor} as
$
  \operatorname{MF}(\mathbb{C}) \triangleq |\{i \in \mathcal{A} \mid i<\tau(\mathbb{C})\}|.
$
Then we can obtain the number of low-weight codewords via the following method:
\begin{itemize}
  \item Step One: Identify the most likely polar cosets that may contain low-weight codewords, similar to the method described in \cite{liAdaptiveSuccessiveCancellation2012}. In this step, transmit an all-zero codeword over a channel with a very high signal-to-noise ratio (SNR) to a large-list SCL decoder. The difference here is that the SCL decoder doesn't need to decode all information bits; it's sufficient when $\operatorname{MF}(\mathbb{C})$ information bits are decoded. Then we obtain $L$ different polar cosets generated by decoding paths with length $\tau(\mathbb{C})$.
  \item Step Two: Utilize {\tt{CalA()}} to calculate the weight enumerating functions for $L$ polar cosets obtained in the first step. The coefficients of low-degree rational terms in the sum of these weight enumerating functions represent the number of low-weight codewords in the polar code or PAC code.
  \item We can continuously increase the list size in the SCL decoder to achieve convergence in the results and obtain the number of codewords with larger weights.
\end{itemize}

The efficiency of this method is closely related to the size of $k-\operatorname{MF}(\mathbb{C})$. The larger the $k-\operatorname{MF}(\mathbb{C})$, the more efficient the method becomes. Therefore, this method may not be suitable for some precoded polar codes like CRC-polar codes with $\operatorname{MF}(\mathbb{C})=k$. In that case, this method is equivalent to the method in \cite{liAdaptiveSuccessiveCancellation2012}.
\subsection{Number of High-Weight Codewords}
\label{sec_3_subsec_2}
High-weight codewords constitute the majority of all codewords in a polar code, and their distribution significantly impacts the performance of polar codes. However, obtaining the accurate weight distribution of high-weight codewords often requires exhaustive enumeration, which leads to exponential computational complexity for long polar codes, making it practically infeasible.

In this section, we continue to leverage the {\tt{CalA()}} function to simulate the weight distribution of high-weight codewords. Within an acceptable error tolerance, the complexity of obtaining the distribution of high-weight codewords is substantially reduced, as Algorithm \ref{alg1} shown.

\begin{algorithm}  
  \label{alg1}
  \caption{Simulate the distribution of high-weight codewords}  
  \LinesNumbered  
  \KwIn{error tolerance $\varepsilon $}  
  \KwOut{the distribution of high-weight codewords $A_{sum}(X)$}
  initialize a sliding filter $\mathbf{D}$ and the distribution of high-weight codewords $A_{sum}(X)=0$;\\
  \While(){$\operatorname{mean}(\mathbf{D}) \geq \varepsilon $}
  {  
      Generate a polar coset generated by a random polar-encoded sequences of length $\tau(\mathbb{C})$, then calculate its weight enumeration function $A_{temp}(X)$ by {\tt{CalcA()}};\\
        $A_{sum}^{\prime}(X)=A_{sum}(X)+A_{temp}(X)$;\\
       Compute the normalized logarithm difference between $A_{sum}^{\prime}(X)$ and $A_{sum}(X)$: $\epsilon = |(\log{A_{sum}}-\log{\sum {A_{sum}}})-(\log{A_{sum}^{\prime}}-\log{\sum {A_{sum}^{\prime}}})|$;($A_{sum}$ and $A_{sum}^{\prime}$ are sequences of coefficients of $A_{sum}(X)$ and $A_{sum}^{\prime}(X)$, respectively)\\
        Push $\epsilon$ into $\mathbf{D}$;\\
        Renew $A_{sum}(X)=A_{sum}^{\prime}(X)$;
  }  
  \textbf{return} $A_{sum}$;
\end{algorithm} 
We will give a brief explanation of why continually accumulating the weight enumeration functions of randomly generated cosets allows us to get progressively closer to the distribution of high-weight codewords:

Let $\mathbb{C}$ denote the polar code which can be partitioned by two subsets $\mathbb{C}\{u_i=0\}$ and $\mathbb{C}\{u_i=1\}$ according to $u_i$ with $0\leq i\leq N-1$. Continuing in this manner, $\mathbb{C}$ can be divided into:
\begin{equation*}
\begin{aligned}
	\mathbb{C} &=\mathbb{C} \left\{ u_{a_0}=0,u_{a_1}=0,\cdots ,u_{a_{\operatorname{MF}(\mathbb{C})-1}}=0,u_{a_{\operatorname{MF}(\mathbb{C})}}=0 \right\}\\
	&\cup \mathbb{C} \left\{ u_{a_0}=0,u_{a_1}=0,\cdots ,u_{a_{\operatorname{MF}(\mathbb{C})-1}}=0,u_{a_{\operatorname{MF}(\mathbb{C})}}=1 \right\}\\
	&\cup \mathbb{C} \left\{ u_{a_0}=0,u_{a_1}=0,\cdots ,u_{a_{\operatorname{MF}(\mathbb{C})-1}}=1,u_{a_{\operatorname{MF}(\mathbb{C})}}=0 \right\}\\
	&\cup \cdots\\
	&\cup \mathbb{C} \left\{ u_{a_0}=1,u_{a_1}=1,\cdots ,u_{a_{\operatorname{MF}(\mathbb{C})-1}}=1,u_{a_{\operatorname{MF}(\mathbb{C})}}=0 \right\}\\
	&\cup \mathbb{C} \left\{ u_{a_0}=1,u_{a_1}=1,\cdots ,u_{a_{\operatorname{MF}(\mathbb{C})-1}}=1,u_{a_{\operatorname{MF}(\mathbb{C})}}=1 \right\}\\
\end{aligned}
\end{equation*}
where $a_j (0\leq j\leq \operatorname{MF}(\mathbb{C}))$ is the $j+1$-th element of $\mathcal{A}$. From \cite{yaoDeterministicAlgorithmComputing2023}, it is evident that based on the properties of lower triangular affine (LTA) groups, many polar cosets share the same weight enumerating functions. Assuming that there are $B$ different weight enumeration functions for all sub-codewords, let's denote these weight enumerating functions as $A_{\mathbb{C}_l}$, where $1 \leq l \leq B$. In this case, the weight enumerating function of polar codes can be expressed as follows:
\begin{equation*}
  \begin{aligned}
    A_{\mathbb{C}}(X)&=\alpha _1A_{\mathbb{C} _1}(X)+\alpha _2A_{\mathbb{C} _1}(X)+
    \cdots \\
    & +\alpha _{B-1}A_{\mathbb{C} _{B-1}}(X)+\alpha _BA_{\mathbb{C} _B}(X)
  \end{aligned}
\end{equation*}
where $A_{\mathbb{C}}(X)$ denotes the weight distribution of the polar code $\mathbb{C}$ and $\alpha_l$ represents the proportion of $A_{\mathbb{C}_l}$ in the overall weight distribution. Therefore, we can approximate the true proportions by randomly generating a certain number of sub-codewords and observing their frequencies. The efficiency of this method is also related to the size of $k-\operatorname{MF}(\mathbb{C})$.

\section{Simulations analysis and discussion}
\label{sec4}

\subsection{Analysis of the Effectiveness of the Method}

\begin{table}[!ht]
  \centering
  \caption{the number of low-weight codewords for (128,64) 5G Polar Code}
  \label{tab_1}

  \begin{tabular}{|c|c|c|c|c|c|}
  \hline
      List Size & $A_8$ & $A_{12}$ & $A_{16}$ & $A_{20}$ & Time \\ \hline
      8 & 296 & 672 & 28456 & 92960 & 0.002s \\ \hline
      16 & 304 & 768 & 30392 & 116992 & 0.006s \\ \hline
      1000 & 304 & 768 & 91598 & 1347016 & 0.256s \\ \hline
      5835 & 304 & 768 & 161528 & 3140176 & 1.536s \\ \hline
      Exact Number & 304 & 768 & 161528 & 4452096 & - \\ \hline
  \end{tabular}
\end{table}

\begin{table}[!ht]
  \centering
  \caption{the number of low-weight codewords for (64,32) RM-constructed PAC Code}
  \label{tab_2}

  \begin{tabular}{|c|c|c|c|c|c|}
  \hline
      List Size & $A_{8}$ & $A_{10}$ & $A_{12}$ & $A_{14}$ & Time \\ \hline
      8 & 176 & 0 & 672 & 0 & 0.003s \\ \hline
      64 & 536 & 176 & 4992 & 1936 & 0.017s \\ \hline
      100 & 536 & 512 & 5472 & 5632 & 0.026s \\ \hline
      1000 & 536 & 512 & 10624 & 24064 & 0.197s \\ \hline
      Exact Number & 536 & 512 & 10624 & 24064 & - \\ \hline
  \end{tabular}
\end{table}

\begin{figure}[!t]
  \centering
  \includegraphics[width=\columnwidth]{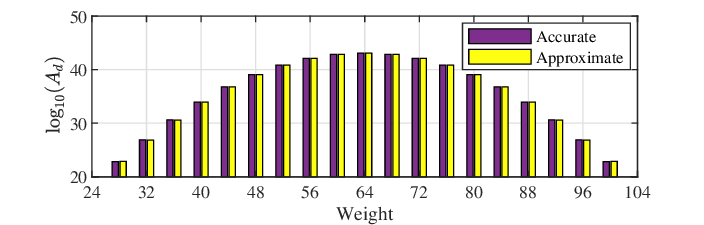}
  \caption{the comparison between the exact
  and approximate distributions of high-weight codewords for
  (128,64) 5G Polar code}
  \label{fig_1}
\end{figure}

\begin{figure}[!t]
  \centering
  \includegraphics[width=\columnwidth]{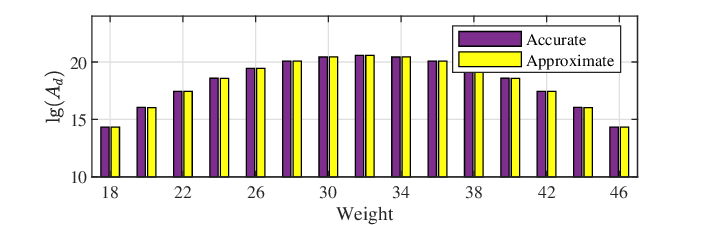}
  \caption{the comparison between the exact
  and approximate distributions of high-weight codewords for
  (64,32) RM-constructed PAC code}
  \label{fig_2}
\end{figure}

Table \ref{tab_1} and \ref{tab_2} provide the number of low-weight codewords of (128,64) 5G polar code and (64,32) Reed-Muller (RM) constructed PAC code\footnote[1]{The reason for not using the (128, 64) RM-constructed PAC code here is that obtaining its precise weight distribution requires an exceptionally large amount of computation.}, respectively. Rows 2 to 4 show the number of low-weight codewords obtained using the method described in Section \ref{sec_3_subsec_1} with different list sizes, while the last row shows the exact number obtained using exhaustive algorithms. Since the \emph{mixing factor} for 5G polar code is relatively small, a small list size and a short time are sufficient to accurately determine the number of low-weight codewords using the proposed method. For other low-weight codewords, we can still increase the list size gradually to obtain them. As RM-constructed codes have the largest mixing factor (as proven in \cite{yaoDeterministicAlgorithmComputing2023}), the efficiency of the proposed method becomes lower but still effective. 

Fig. \ref{fig_1} and Fig. \ref{fig_2} illustrate the comparison between the exact and approximate distributions of high-weight codewords for (128,64) 5G polar code and (64,32) RM constructed PAC code, respectively. The error tolerance for the sliding filter is set to $10^{-5}$. Using Algorithm \ref{alg1}, we obtained the approximate distribution of high-weight codewords in less than a minute. From the figures, it can be observed that the approximate distribution is nearly identical to the exact distribution. Specifically, for (128,64) 5G polar code, the exact count of codewords with $d=N/2$ is $e^{43.0941}$, while the estimated value is $e^{43.0944}$, resulting in a relative error of $|e^{43.0941}-e^{43.0944}|/e^{43.0941}=0.03\%$. And for (64,32) PAC code, the relative error is $0.44\%$.

\subsection{Analysis of the Distinctions Between PAC code and Polar code}
\begin{table}[!ht]
  \centering
  \caption{the number of low-weight codewords obtained using the method described in Section III with different list sizes}
  \label{tab_3}
  \resizebox{\columnwidth}{!}{
  \begin{tabular}{|c||c|c|}
  \hline
      Code & $d_{\min}$ & Number of Low-Weight Codewords  \\ \hline
      (128,64) 5G Polar & 8 & $A_8=304$, $A_{12}=768$, $A_{16}=161528$ \\ \hline
      (128,64) RM PAC & 16 &$A_{16}=26392$, $A_{18}=13056$, $A_{20}=223232$\\ \hline
      (128,64) GA PAC & 8 & $A_8=256$, $A_{12}=960$, $A_{16}=106264$\\ \hline
      (256,128) 5G Polar & 8 &$A_8=32$, $A_{16}=93296$, $A_{20}=90112$  \\ \hline
      (256,128) RM PAC & 16 &$A_{16}=25168$, $A_{18}=3008$, $A_{20}=138688$ \\ \hline
      (256,128) GA PAC & 8 &$A_8=32$, $A_{16}=51568$, $A_{18}=9472$\\ \hline
      (512,256) 5G Polar & 16 &$A_{16}=52832$, $A_{24}=18198016$, $A_{28}\geq 33295296$  \\ \hline
      (512,256) RM PAC & 32 &$A_{32}=1032243$, $A_{40}=113936$, $A_{48}\geq8035564$ \\ \hline
      (512,256) GA PAC & 16 &$A_{16}=41824$, $A_{18}=832$, $A_{20}\geq44800$\\ \hline      
  \end{tabular}
  }
\end{table}

\begin{figure}[!t]
  \centering
  \includegraphics[width=\columnwidth]{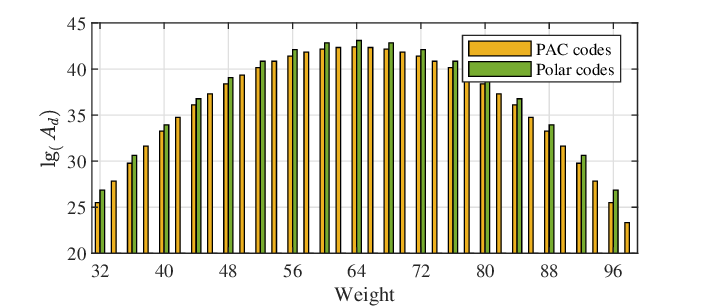}
  \caption{the comparison of the distribution of high-weight codewords for (128,64) 5G Polar code and (128,64) RM-constructed PAC code}
  \label{fig_3}
\end{figure}

Table \ref{tab_3} presents the number of low-weight codewords for PAC codes and polar codes with different code lengths and constructions where GA stands for Gaussian Approximation. Fig. \ref{fig_3} provides a comparison of the distribution of high-weight codewords for (128,64) 5G polar code and (128,64) RM-constructed PAC code. By examining Table \ref{tab_3} and Fig. \ref{fig_3}, we can intuitively observe how PAC codes have improved upon polar codes in terms of weight distribution. The RM construction for PAC codes notably increases the minimum weight compared to polar codes, resulting in a significant enhancement in BLER performance, especially for short code lengths. Additionally, the pre-convolution operation introduced by PAC codes effectively reduces the number of low-weight codewords while increasing the types of medium-weight codewords, leading to a smoother weight distribution. Both of these aspects contribute to enhancing the performance of polar codes. 

The comparative results in this part also provide an additional insight into why PAC codes constructed by GA exhibit lower computational complexity in the Fano algorithm. This is attributed to their smaller minimum distance, which makes it easier to retrace the correct path during the backtracking process.

\subsection{Discussion}
Compared to the exact weight distribution of a (128, 64) 5G polar code, which took approximately two hours to obtain in \cite{yaoDeterministicAlgorithmComputing2023}, our method can provide an extremely accurate approximate distribution within just one minute. For longer codes, \cite{miloslavskayaComputingPartialWeight2022} requires hours of computation, while our approach can yield reasonably satisfactory results within just an hour.

Once we have obtained an approximate weight distribution
of the code, we can predict the code’s BLER performance
through the union bound, aiding in the optimization of polar codes
and PAC codes. Concerning PAC codes, additional optimization efforts may target the reduction of low-weight codewords through appropriate construction methods or the ongoing enhancement of their minimum Hamming distance. Some research on the construction and performance of polar codes in fading channels also relies on the
code’s weight distribution. One can also investigate the optimization and performance of PAC codes in fading channels from the perspective of weight distribution.

\section{Conclusion}
\label{sec5}

In this letter, we introduce an efficient method to estimate the weight distribution for polar codes and PAC codes. For low-weight codewords, a series of polar cosets possibly containing low-weight codewords are collected through an SCL decoder. Subsequently, a deterministic recursive algorithm is applied to calculate the sum of weight distributions of these polar cosets, thereby determining the number of low-weight codewords. Regarding high-weight codewords, the method estimates the weight distribution by a subset of randomly generated polar cosets. Simulation results demonstrate that this hybrid method maintains low computational complexity while accurately determining the number of low-weight codewords and providing an approximate distribution of high-weight codewords.

\ifCLASSOPTIONcaptionsoff
  \newpage
\fi



%

\bibliographystyle{IEEEtran}
\bibliography{IEEEabrv,cite}

%








\end{document}